\documentclass[10pt,english]{revtex4-1}
\usepackage{mathptmx}
\usepackage[T1]{fontenc}
\usepackage[utf8]{luainputenc}
\usepackage{babel}
\usepackage{amssymb}
\usepackage{graphicx}
\usepackage[unicode=true,pdfusetitle,
 bookmarks=true,bookmarksnumbered=false,bookmarksopen=false,
 breaklinks=false,pdfborder={0 0 1},backref=false,colorlinks=false]
 {hyperref}

\makeatletter

\newcommand{\lyxdot}{.}

\makeatother

\begin{document}

\title{Energy of the Interacting Self-Avoiding Walk at the $\theta-$point}

\author{Simone Franchini and Riccardo Balzan}

\affiliation{Dipartimento di Fisica, Sapienza Università di Roma, P.le Aldo Moro
1, 00185, Roma, Italy}
\begin{abstract}
We perform a numerical study of a new microcanonical polymer model
on the three dimensional cubic lattice, consisting of ideal chains
whose range and number of nearest-neighbor contacts are fixed to given
values. Our simulations suggest an interesting exact relation concerning
the internal energy per monomer of the Interacting Self-Avoiding Walk
at the $\theta-$point.

~
\end{abstract}
\maketitle

\section{Introduction}

It is well known that a polymer chain can collapse from an extended
to a compact configuration if the temperature or the solvent quality
is lowered below some critical value. This phenomenon, known as Coil-to-Globule
transition (CG, \cite{Flory,CG,CGDNA}), arises when the attractive
interaction between the monomers overwhelm the excluded volume effect.
At the transition temperature (commonly called $\theta-$point) these
contributions compensate, resulting in a phase where the chains behave
approximately as random walks \cite{Flory,De Gennes-Scal.,des Cloizeaux,Grosberg-Kuznestov}. 

Let $\omega_{N}$ be an $N-$steps Simple Random Walk (SRW) on the
cubic lattice $\mathbb{Z}^{d}$, 
\begin{equation}
\omega_{N}=\left\{ x_{t}\left(\omega_{N}\right)\in\mathbb{Z}^{d}:\,0\leq t\le N\right\} ,
\end{equation}
by convention we fix the seed monomer at $x_{0}\left(\omega_{N}\right)=0$.
The chain can be represented trough the locations of its monomers
$x_{t}\left(\omega_{N}\right)$ or equivalently by the orientations
of its steps
\begin{equation}
x_{t}\left(\omega_{N}\right)-x_{t-1}\left(\omega_{N}\right)\in\Omega_{1},
\end{equation}
where $\Omega_{1}$ is the set of possible orientations on $\mathbb{Z}^{d}$
(for the cubic lattice the number is $|\Omega_{1}|=2d$). Then, we
indicate with 
\begin{equation}
\Omega_{N}=\Omega_{1}^{N}\ni\omega_{N}
\end{equation}
the set of all possible chain configurations.

\begin{figure}
~

~

\includegraphics[scale=0.28]{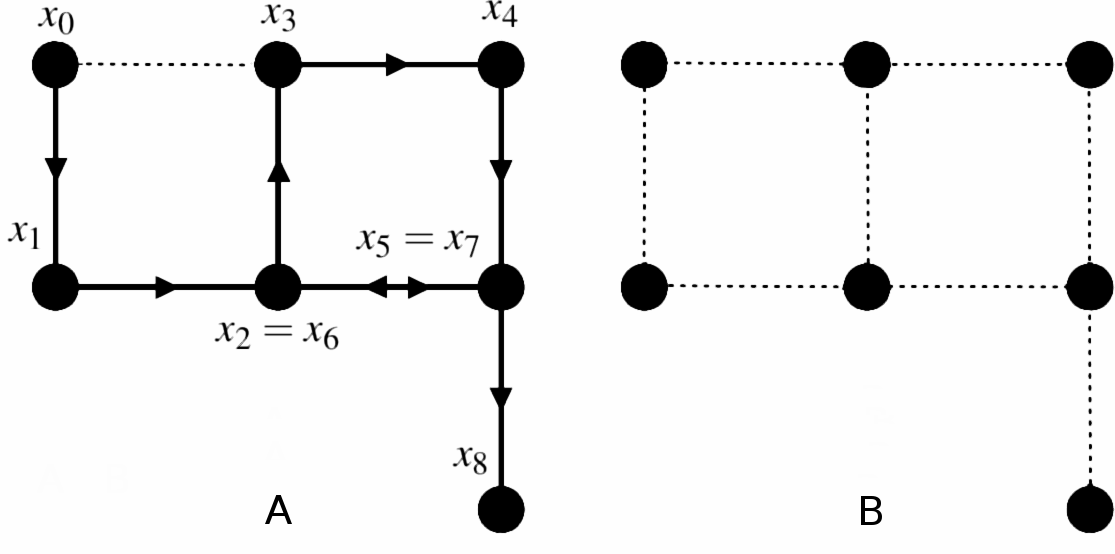}

~

\caption{\label{FIG1}Range and link count for a chain $\omega_{8}=\left\{ x_{0},x_{1},\,...\,,x_{8}\right\} $
of $N=8$ steps on $\mathbb{Z}^{2}$, shown on top. A shows the actual
walk, while B highlights the range points (black circles) and the
links (dotted segments) of $\omega_{8}$. The total range is $R\left(\omega_{8}\right)=7$,
the number of self-intersections is then $M\left(\omega_{8}\right)=8+1-R\left(\omega_{8}\right)=2$,
occurring at the $6-$th and $7-$th steps. The total number of links
is $L\left(\omega_{8}\right)=8$, as it counts also the links imposed
by the chain condition (in A the only non-trivial link is that between
monomers $x_{0}$ and $x_{3}$). }
\end{figure}

Here we present a micro-canonical model where the number of distinct
lattice sites visited by the walk $R\left(\omega_{N}\right)$ (\textit{range})
and the number of nearest-neighbors monomer pairs $L\left(\omega_{N}\right)$
(\textit{links}) are constrained to scale with the number of steps
$N$, formally 
\begin{equation}
R\left(\omega_{N}\right)=\left\lfloor \left(1-m\right)N\right\rfloor ,\ L\left(\omega_{N}\right)=\left\lfloor \lambda N\right\rfloor ,
\end{equation}
where we denoted by $\left\lfloor z\right\rfloor $ the lower integer
truncation of $z\in\mathbb{R}$, (see Figure \ref{FIG1}). The model
is controlled by the pair of parameters $m$ and $\lambda$, and the
Interacting Self-Avoiding Walk (ISAW, \cite{ISAW1,ISAW2,ISAW3,ISAW4,ISAW5,ISAW6,Douglas,Madras-Slade})
is recovered by taking $m=0$. 

We numerically investigated the micro-canonical phase diagram on the
plane $\left(m,\lambda\right)$, formulating a conjecture on the location
of the transition line $\lambda=\ell_{c}\left(m\right)$ that is expected
to separate the SAW like-phase (where the scaling of the average chain
displacement is that of the SAW) from the clustered phase (in which
the chains configure into compact clusters). 

Based on these computer simulations and some additional theoretical
arguments, our analysis suggests that at least in the Thermodynamic
Limit (TL) $N\rightarrow\infty$ the critical link density is a linear
function of $m$ 
\begin{equation}
\ell_{c}\left(m\right)=\lambda_{c}+\delta_{c}m
\end{equation}
and the constant $\lambda_{c}$ is expected to match the density of
contacts per monomer of the ISAW at the $\theta-$point in the TL.

Before going further we introduce the notation and state some basic
properties. Without loss of generality, instead of $R\left(\omega_{N}\right)$
we will work with the related quantity
\begin{equation}
M\left(\omega_{N}\right)=N+1-R\left(\omega_{N}\right),
\end{equation}
which represent the number of \textit{intersections} present in the
chain $\omega_{N}$. Our model is then defined by a partition of $\Omega_{N}$
into subsets $\Omega_{N}\left(M,L\right)$ such that each walk has
exactly $M$ intersections and $L$ links 
\begin{equation}
\Omega_{N}\left(M,L\right)=\left\{ \omega_{N}\in\Omega_{N}:\,M\left(\omega_{N}\right)=M,\,L\left(\omega_{N}\right)=L\right\} ,
\end{equation}
we indicate with the symbol $\langle\,\cdot\,\rangle_{M,L}$ the average
at fixed $N$, $M$ and $L$ 
\begin{equation}
\langle\,\cdot\,\rangle_{M,L}=\frac{1}{\left|\Omega_{N}\left(M,L\right)\right|}\sum_{\omega_{N}\in\Omega_{N}\left(M,L\right)}\left(\,\cdot\,\right),
\end{equation}
while the dependence on $N$ is kept implicit. Also, we can define
the probability of uniformly extracting a chain with $M$ intersections
and $L$ links 
\begin{equation}
p_{0}\left(M,L\right)=\frac{\left|\Omega_{N}\left(M,L\right)\right|}{\left(2d\right)^{N}}
\end{equation}
that by definition sums to one
\begin{equation}
\sum_{M,L}p_{0}\left(M,L\right)=1.
\end{equation}
We remark that the link counter $L\left(\omega_{N}\right)$ also includes
the links between consecutive monomers, hence is always bounded by
the range $R$ from below and by $dR$ from above, for $d=3$ 
\begin{equation}
1\leq\frac{L\left(\omega_{N}\right)}{N+1-M\left(\omega_{N}\right)}<3,
\end{equation}
also, notice that $L\left(\omega_{N}\right)$ can increase only if
$M\left(\omega_{N}\right)$ does not (the variables are anti-correlated).

In the simplest case, the CG transition can be modeled by incorporating
attractive nearest-neighbors interactions in the Self-Avoiding Walk
(SAW) \cite{DJMODEL,DJMODEL2,Slade RG Theory,Madras-Slade,Huges,Slade},
the canonical version of our model is described by the Hamiltonian
\begin{equation}
H\left(\omega_{N}\right)=\epsilon M\left(\omega_{N}\right)+\gamma L\left(\omega_{N}\right).
\end{equation}
The competition between the repulsive range term $\epsilon M\left(\omega_{N}\right)$
versus the attractive nearest-neighbor interaction $\gamma L\left(\omega_{N}\right)$
allows for the CG transition. 

Given the parameters $\beta_{1}/\beta=\epsilon$ and $\beta_{2}/\beta=\gamma$
the associated Gibbs measure is 
\begin{equation}
\mu_{\beta}\left(\omega_{N}\right)=\frac{e^{-\beta_{1}M\left(\omega_{N}\right)-\beta_{2}L\left(\omega_{N}\right)}}{Z_{\beta}}
\end{equation}
Notice that the partition function can be expressed as a sum over
$M$ and $L$ using the formula
\begin{equation}
Z_{\beta}=\sum_{\omega_{N}\in\Omega_{N}}e^{-\beta_{1}M\left(\omega_{N}\right)-\beta_{2}L\left(\omega_{N}\right)}=\sum_{M,L}\left|\Omega_{N}\left(M,L\right)\right|e^{-\beta_{1}M-\beta_{2}L},\label{eq:TOTAL}
\end{equation}
and we can also define a pseudo-Gibbs measure
\begin{equation}
p_{\beta}\left(M,L\right)=\frac{\left|\Omega_{N}\left(M,L\right)\right|e^{-\beta_{1}M-\beta_{2}L}}{Z_{\beta}}
\end{equation}
that allows to express the thermal averages 
\begin{equation}
\langle\,\cdot\,\rangle_{\beta}=\sum_{\omega_{N}\in\Omega_{N}}\mu_{\beta}\left(\omega_{N}\right)\left(\,\cdot\,\right)=\sum_{M,L}p_{\beta}\left(M,L\right)\langle\,\cdot\,\rangle_{M,L}
\end{equation}
in terms of the microcanonical averages $\langle\,\cdot\,\rangle_{M,L}$. 

Based on the existing literature on the IDJ model \cite{Madras-Slade,Clisby,Clisby2}
, the limit $N\rightarrow\infty$ of our model should exist for any
choice of the parameters, and then we expect that for any $\beta$
and any ratio $\beta_{1}/\beta_{2}$ the probability measure $p_{\beta}(\left\lfloor mN\right\rfloor ,\left\lfloor \lambda N\right\rfloor )$
concentrates on some point of the $(m,\lambda)$ plane. 

We indicate with $M_{N}$ the average number of intersections for
a SRW of $N$ steps 
\begin{equation}
M_{N}=\sum_{M,L}p_{0}\left(M,L\right)\cdot M,
\end{equation}
while $L_{N}$ is the average number of links
\begin{equation}
L_{N}=\sum_{M,L}p_{0}\left(M,L\right)\cdot L,
\end{equation}
By standard SRW theory \cite{Douglas,Spitzer,Huges,Feller}, the average
densities of intersections and links are given by the formulas 
\begin{equation}
M_{N}=m_{0}N+u_{0}\sqrt{N}+o\left(\sqrt{N}\right),
\end{equation}
\begin{equation}
L_{N}=\lambda_{0}N+w_{0}\sqrt{N}+o\left(\sqrt{N}\right),
\end{equation}
the constants can be exactly computed (for example $m_{0}=C_{3}$
Polya constant \cite{Franchini}). Also, the fluctuations 
\begin{equation}
\Delta M\left(\omega_{N}\right)=M\left(\omega_{N}\right)-M_{N},
\end{equation}
\begin{equation}
\Delta L\left(\omega_{N}\right)=L\left(\omega_{N}\right)-L_{N},
\end{equation}
are expected to satisfy a joint Central Limit Theorem (CLT) centered
at zero, and $p_{0}\left(M,\,L\right)$ should concentrate in a $O\left(\sqrt{N}\right)$
neighborhood of the point $\left(m_{0}N,\lambda_{0}N\right)$ on the
$\left(M,L\right)$ space. As we shall see in short, this fact is
of central importance to locate the critical line in three dimensions.
We will discuss its grounds when dealing with the conjectured phase
diagram.

\section{Locating the transition line}

It is easy to verify that the proposed Hamiltonian converges to the
ISAW in the limit $\epsilon\rightarrow\infty$ (if also $\beta_{2}=0$
corresponds to the SAW). Under the assumption that $\log p_{0}\left(0,\left\lfloor \lambda N\right\rfloor \right)$
is convex in $\lambda$ at least in the SAW phase, we can expect that
\begin{equation}
\lim_{N\rightarrow\infty}\lim_{\beta_{1}\rightarrow\infty}\lim_{\beta_{2}\rightarrow\beta_{c}}\frac{\langle L\left(\omega_{N}\right)\rangle_{\beta}}{N}=\lim_{N\rightarrow\infty}\frac{\langle L\left(\omega_{N}\right)\rangle_{0,\left\lfloor \lambda_{c}N\right\rfloor }}{N}=\lambda_{c},\label{eq:eree}
\end{equation}
ie, that in the TL the critical energy densities should be the same
in both the canonical and microcanonical versions. 
\begin{figure}
\includegraphics[scale=0.26]{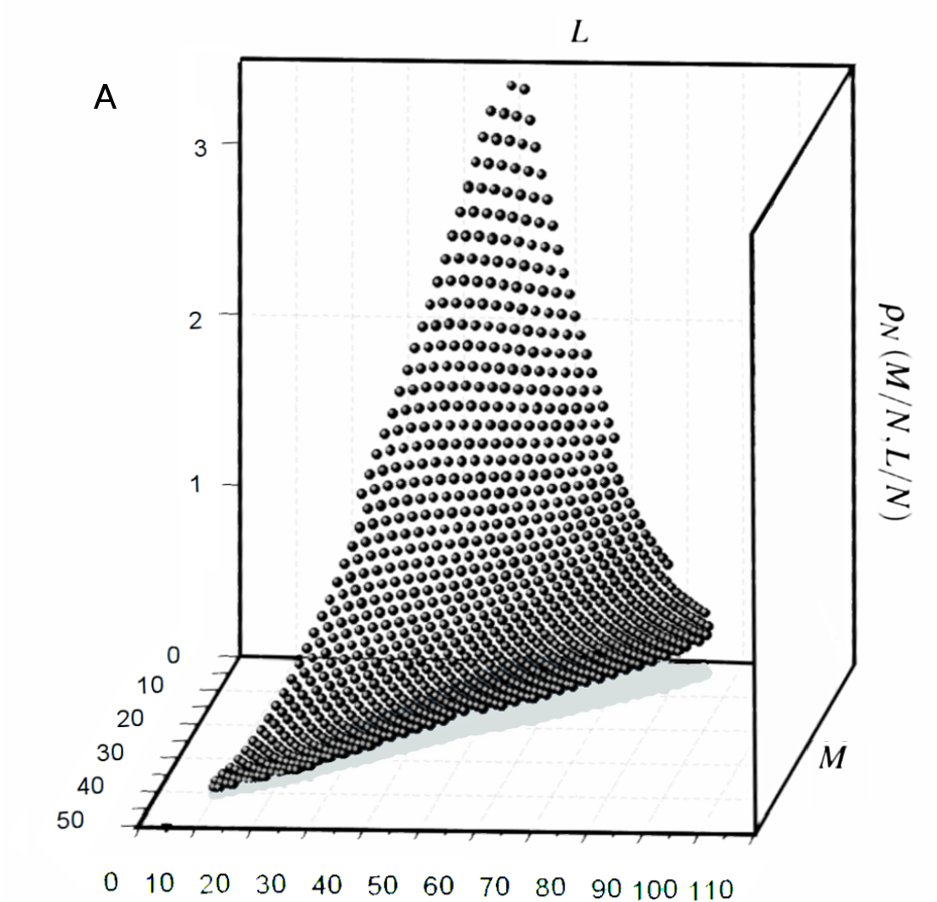}

~

\includegraphics[scale=0.24]{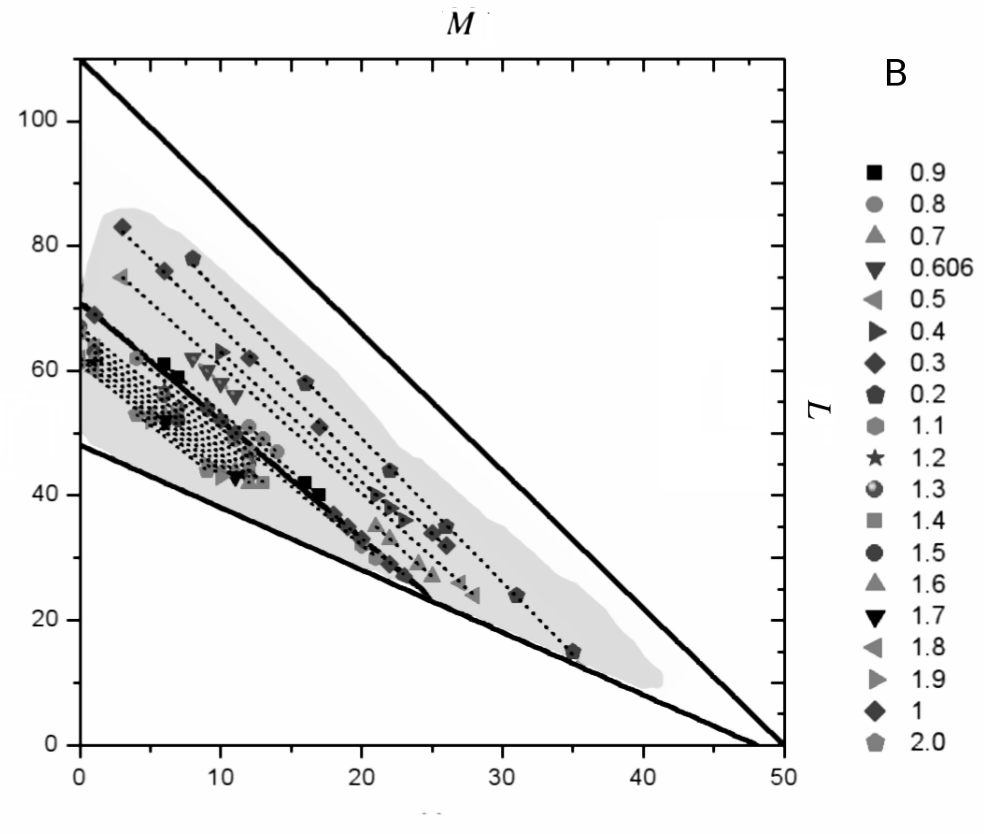}

\caption{\label{FIG2}Surface $\rho_{N}\left(m,\lambda\right)$ for a ISAW
of $N=50$. In figure A the surface $\rho_{N}\left(m,\lambda\right)$
is computed for a large part of the parameter space using a PERM algorithm
(gray area in B). Figure B shows some level lines $\rho_{N}\left(m,\lambda\right)=r$
as scatter points, the line $\rho_{N}\left(m,\lambda\right)=1$ and
the boundaries of the allowed parameter space are highlighted by solid
lines. Although the considered chains are very small, the linear behavior
of the level lines in B is still surprising. A simulation of a larger
chain of $N=100$ steps (not shown) gave the same picture.}
\end{figure}

To present the essential features of the phase diagram we will first
discuss the quantity 
\begin{equation}
\nu\left(m,\lambda\right)=\lim_{N\rightarrow\infty}\frac{\log\,\langle x_{N}^{2}\left(\omega_{N}\right)\rangle_{\left\lfloor mN\right\rfloor ,\left\lfloor \lambda N\right\rfloor }}{2\log N},
\end{equation}
which represents the critical exponent of the squared end-to-end distance
when $M$ and $L$ are constrained to grow proportionally to $N$. 

For $\gamma\rightarrow0$ we obtain the so called Stanley model for
$\epsilon>0$, of Hamiltonian $H_{0}\left(\omega_{N}\right)=\epsilon M\left(\omega_{N}\right)$,
while for $\epsilon<0$ is the Rosenstock Trapping model. The corresponding
microcanonical model is 
\begin{equation}
\Omega_{N}\left(M\right)=\bigcup_{L}\Omega_{N}\left(M,L\right)
\end{equation}
and has been studied in \cite{Franchini,frabalz} where numerical
simulations and additional theoretical arguments support the conjecture
that the displacement exponent of the set $\Omega_{N}\left(\left\lfloor mN\right\rfloor \right)$
\begin{equation}
\nu\left(m\right)=\lim_{N\rightarrow\infty}\frac{\log\,\langle x_{N}^{2}\left(\omega_{N}\right)\rangle_{\left\lfloor mN\right\rfloor }}{2\log N},
\end{equation}
has a drop around $m_{c}=C_{3}$, with a drop band slowly narrowing
as $O\left(1/N^{\alpha}\right)$ and $\alpha=0.29\pm0.1$ (\cite{Franchini},
an independent scaling analysis, not shown, gave $0.31\pm0.1$).

Based on these preliminary studies we conjecture that for any value
of $m$ there is some critical link density $\ell_{c}\left(m\right)$
such that if $\lambda<\ell_{c}\left(m\right)$ the exponent $\nu\left(m,\lambda\right)$
matches the critical exponent $\nu_{3}$ of the SAW. The conjectured
phase diagram is then 
\begin{equation}
\nu\left(m,\lambda\right)=\left\{ \begin{array}{ccc}
\nu_{3} &  & \lambda<\ell_{c}\left(m\right)\\
1/2 &  & \lambda=\ell_{c}\left(m\right)\\
1/3 &  & \lambda>\ell_{c}\left(m\right)
\end{array}\right.
\end{equation}
where $\nu_{3}$ is the critical exponent of the SAW governing the
end-to-end distance \cite{Madras-Slade,Clisby,Clisby2}. If the link
density is exactly $\lambda=\ell_{c}\left(m\right)$ the energy contributions
from range and links should balance, giving a SRW-like critical behavior
with exponent $\nu\left(m,\ell_{c}\left(m\right)\right)=1/2$, while
for $\lambda>\ell_{c}\left(m\right)$ we expect to be in the cluster
phase, then $\nu\left(m,\lambda\right)=1/3$. Notice that for $m\rightarrow0$
we must have $\ell_{c}\left(0\right)=\lambda_{c}$ energy density
of the ISAW at the theta point. 

\begin{figure}
\includegraphics[scale=0.21]{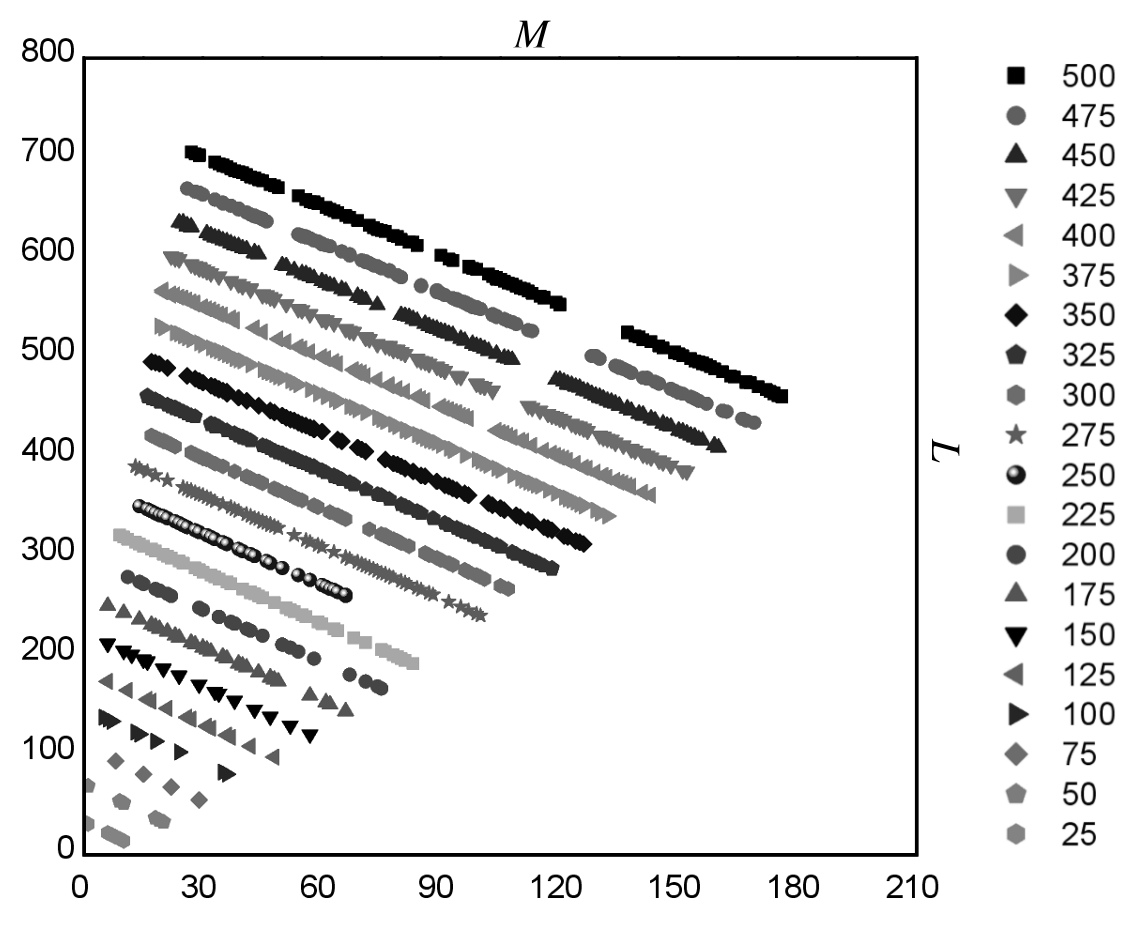}

\caption{\label{FIG6}Transition line $\rho_{N}\left(m,\lambda\right)=1$ for
chains up to $N=500$ for a large portion of the parameter space using
a PERM algorithm. The line from different $N$ are shown on the same
graph to allow comparison. The lenght of the chains varies from $N=25$
to $500$. The linear behavior of the critical level line seems present
also for longer chains. The intercepts at $M=0$, extrapolated from
linear fits, are shown as white squares in Figure \ref{FIG6-1-1}
A.}
\end{figure}

Although an investigation of the parameter $\nu\left(m,\lambda\right)$
should be carried on to verify the phase exponents (as is done in
\cite{Franchini} for the Range Problem), we believe that the existing
literature on IDJ-like models \cite{De Gennes-Scal.,des Cloizeaux,Grosberg-Kuznestov,ISAW1,ISAW2,ISAW3,ISAW5,ISAW6,DJMODEL,Slade RG Theory,Douglas,Madras-Slade}
already support the existence of a non-trivial transition line, and
we decided to locate $\ell_{c}\left(m\right)$ by computing the level
lines of the estimator
\begin{equation}
\rho_{N}\left(m,\lambda\right)=\frac{\langle x_{N}^{2}\left(\omega_{N}\right)\rangle_{\left\lfloor mN\right\rfloor ,\left\lfloor \lambda N\right\rfloor }}{N},
\end{equation}
that by previous considerations satisfy
\begin{equation}
\rho_{N}\left(m,\lambda\right)=\left\{ \begin{array}{ccc}
O\left(N^{2\nu_{3}-1}\right) &  & \lambda<\ell_{c}\left(m\right)\\
O\left(1\right) &  & \lambda=\ell_{c}\left(m\right)\\
O\left(N^{-2/3}\right) &  & \lambda>\ell_{c}\left(m\right)
\end{array}\right.\label{eq:behavior}
\end{equation}
We computed the set $\ell_{N}\left(m,r\right)$ that satisfy 
\begin{equation}
\rho_{N}\left(m,\ell_{N}\left(m,r\right)\right)=r
\end{equation}
by numerical simulations using a PERM algorithm \cite{PERM,Grassberger,Prellberg,Hsu-Grassberger}.
For very short chains $\left(N\leq100\right)$ we were able to explore
a large portion of the space $\left(m,\lambda\right)$, with $r$
ranging from small values up to the scale of $\rho_{N}\left(0,1\right).$
We found that for very short chains
\begin{equation}
N\ell_{N}\left(m,r\right)=\left\lfloor \lambda_{N}\left(r\right)N+\delta_{N}\left(r\right)\cdot mN\right\rfloor 
\end{equation}
is verified with extremely high accuracy at any observed $r$. For
small chains we observe that the level curves of $\rho_{N}\left(m,\lambda\right)$
appears to be straight lines (see Figure \ref{FIG6}). 

Given the small size of the chains we cannot conclude much from this
observation, but driven by this preliminary experiment we decided
to fix $r=1$, that is the diffusion behavior of the SRW, and perform
an intensive investigation of the curve $\ell_{N}\left(m,1\right)$,
\begin{equation}
\rho_{N}\left(m,\ell_{N}\left(m,1\right)\right)=1,
\end{equation}
that by Eq.(\ref{eq:behavior}) is expected to converge to the critical
line in the thermodynamic limit \cite{levelchoice} 
\begin{equation}
\lim_{N\rightarrow\infty}\ell_{N}\left(m,1\right)=\ell_{c}\left(m\right).
\end{equation}

The PERM algorithm, which is very efficient in simulating $\theta-$point
chains, allowed to evaluate $\ell_{N}\left(m,1\right)$ up to chains
with $N=500$ in a macroscopic portion of the $\left(M,L\right)$
space. We found stronger evidences that at least the curve $\ell_{N}\left(m,1\right)$
is still a line up to integer truncation, 
\begin{equation}
N\ell_{N}\left(m,1\right)=\left\lfloor \lambda_{N}\left(1\right)N+\delta_{N}\left(1\right)\cdot mN\right\rfloor ,
\end{equation}
suggesting the conjecture that the critical line may remain a line
in the thermodynamic limit, with critical coefficients eventually
satisfying
\begin{equation}
\lim_{N\rightarrow\infty}\lambda_{N}\left(1\right)=\lambda_{c},\ \lim_{N\rightarrow\infty}\delta_{N}\left(1\right)=\delta_{c}.
\end{equation}

This property can be explained as follows. As in \cite{frabalz},
let us partition the chain $\omega_{N}$ into a number $n$ of sub-chains
\begin{equation}
\omega_{N}=\left\{ \omega_{T}^{0},\omega_{T}^{1},\,...\,,\omega_{T}^{n}\right\} 
\end{equation}
each of size $T=N/n$. The sub-chains are indicated with 
\begin{equation}
\omega_{T}^{i}=\left\{ x_{0}^{i},\,x_{1}^{i},\,...\,,x_{T}^{i}\right\} \subset\omega_{N}
\end{equation}
and satisfy the chain constraint $x_{T}^{i}=x_{0}^{i+1}$. If we neglect
the self-intersections between the blocks, as is expected in a SRW-like
chain \cite{Madras-Slade}, we can approximate the probability measure
conditioned on the transition line with a product measure. 

Now, as in \cite{frabalz} we assume that each sub-chain can be either
a critical ISAW, with local densities $\left(0,\lambda_{0}\right)$,
or a SRW, with average local densities $\left(m_{0},\lambda_{0}\right)$.
Then we could write 
\begin{equation}
p_{0}\left(\left\lfloor mN\right\rfloor ,\left\lfloor \ell_{c}\left(m\right)N\right\rfloor \right)\simeq\prod_{i=1}^{n}\,p_{0}\left(0,\left\lfloor \lambda_{c}N\right\rfloor \right)^{\varphi_{i}T}p_{0}\left(\left\lfloor m_{0}N\right\rfloor ,\left\lfloor \lambda_{0}N\right\rfloor \right)^{\left(1-\varphi_{i}\right)T}\label{eq:grgr}
\end{equation}
with $\varphi^{i}\in\left\{ 0,1\right\} $ keeping record of the subchain
type. One in the end finds that under the above product measure condition
the averages of $M\left(\omega_{T}\right)$ and $L\left(\omega_{T}\right)$
satisfy the relation 
\begin{equation}
\langle L\left(\omega_{N}\right)\rangle_{\left\lfloor mN\right\rfloor ,\left\lfloor \ell_{c}\left(m\right)N\right\rfloor }\simeq\lambda_{c}N-\left(\frac{\lambda_{c}-\lambda_{0}}{m_{0}}\right)\langle M\left(\omega_{N}\right)\rangle_{\left\lfloor mN\right\rfloor ,\left\lfloor \ell_{c}\left(m\right)N\right\rfloor }.\label{eq:grgr-1}
\end{equation}

Notice that three dimensional $\theta-$polymers should include logarithmic
corrections to the simple mean-field factorization \cite{De Gennes-Scal.}.
Even if these corrections are important in the usual Range Problem
\cite{frabalz,logcorr} here the constraint to stay on the transition
line forces the chains to behave like SRWs, and we are persuaded that
neglecting these correlations should not affect the shape of the line
in the thermodynamic limit.

\section{A consequence from SRW theory}

An important consequence of the previous conjecture is that the critical
energy density of the ISAW $\lambda_{c}$ would be computable in terms
of SRW measurable quantities.

In fact, we remark that the $p_{0}\left(\left\lfloor mN\right\rfloor ,\left\lfloor \lambda N\right\rfloor \right)$
is expected to concentrate on $\left(m_{0},\lambda_{0}\right)$. Since
the average squared end-to-end distance in the SRW is exactly $N$
we can conclude that also this point must lie on the transition line
\begin{equation}
\ell_{c}\left(m_{0}\right)=\lambda_{0}
\end{equation}
Then, by the previous linearity conjecture we should be able to conclude
that the ratio 
\begin{equation}
\delta_{N}^{*}=\frac{\langle L\left(\omega_{N}\right)\rangle_{\beta}-\langle L\left(\omega_{N}\right)\rangle_{0}}{\langle M\left(\omega_{N}\right)\rangle_{\beta}-\langle M\left(\omega_{N}\right)\rangle_{0}}
\end{equation}
converges to the actual $\delta_{N}$ (and then to the angular coefficient
of the critical line in the TL) under the constraint of constant end-to-end
distance
\begin{equation}
\langle x_{N}^{2}\left(\omega_{N}\right)\rangle_{\beta}=\langle x_{N}^{2}\left(\omega_{N}\right)\rangle_{0}.
\end{equation}
To compute this estimator we expand the Boltzmann factor in the limit
of infinite temperature, ie for small $\beta$ 
\begin{equation}
e^{-\beta_{1}M-\beta_{2}L}=1-\beta_{1}M-\beta_{2}L+O\left(\beta^{2}\right)
\end{equation}
and then compute the averages. It can be shown after some algebra
that in the limit of infinite temperature the differences are approximated
by the expressions
\begin{equation}
\begin{array}{c}
\langle L\left(\omega_{N}\right)\rangle_{\beta}-\langle L\left(\omega_{N}\right)\rangle_{0}=-\beta_{2}\Delta L_{N}^{2}-\beta_{1}\Delta Q_{N}\\
\\
\langle M\left(\omega_{N}\right)\rangle_{\beta}-\langle M\left(\omega_{N}\right)\rangle_{0}=-\beta_{2}\Delta Q_{N}-\beta_{1}\Delta M_{N}^{2}
\end{array}
\end{equation}
where in order to simplify the formulas we introduced a notation for
the variances of links and intersections, 
\begin{equation}
\Delta L_{N}^{2}=\langle\Delta L^{2}\left(\omega_{N}\right)\rangle_{0},\,\Delta M_{N}^{2}=\langle\Delta M^{2}\left(\omega_{N}\right)\rangle_{0},
\end{equation}
and one for the the correlations between $M\left(\omega_{N}\right)$
and $L\left(\omega_{N}\right)$ under the SRW measure
\begin{equation}
\Delta Q_{N}=\langle\Delta M\left(\omega_{N}\right)\Delta L\left(\omega_{N}\right)\rangle_{0}.
\end{equation}
The ratio $\beta_{1}/\beta_{2}$ is obtained from the constraint of
having a constant average end-to-end distance applied to the first
order expansion in $\beta$, 
\begin{equation}
\langle x_{N}^{2}\left(\omega_{N}\right)\rangle_{\beta}-\langle x_{N}^{2}\left(\omega_{N}\right)\rangle_{0}\simeq-\beta_{1}\Delta P_{N}-\beta_{2}\Delta T_{N}=0
\end{equation}
where we again simplified the notation by introducing a symbol for
the correlation between $M\left(\omega_{N}\right)$ and $x_{N}^{2}\left(\omega_{N}\right)$,
\begin{equation}
\Delta P_{N}=\langle\Delta M\left(\omega_{N}\right)\Delta x_{N}^{2}\left(\omega_{N}\right)\rangle_{0}
\end{equation}
and another symbol for the correlation between $L\left(\omega_{N}\right)$
and $x_{N}^{2}\left(\omega_{N}\right)$, which is 
\begin{equation}
\Delta T_{N}=\langle\Delta L\left(\omega_{N}\right)\Delta x_{N}^{2}\left(\omega_{N}\right)\rangle_{0}.
\end{equation}
Finally, substituting the ratio $\beta_{2}/\beta_{1}$ obtained from
the last formula into the approximate expression for $\delta_{N}^{*}$
we obtain the relation 
\begin{equation}
\delta_{N}^{*}=\frac{\Delta Q_{N}+\left(\frac{\Delta P_{N}}{\Delta T_{N}}\right)\Delta L_{N}^{2}}{\Delta M_{N}^{2}+\left(\frac{\Delta P_{N}}{\Delta T_{N}}\right)\Delta Q_{N}}
\end{equation}
that, assuming true our conjecture, would allow to compute the critical
energy density of the ISAW in the TL from the formula 
\begin{equation}
\lambda_{N}^{*}N=L_{N}+\delta_{N}^{*}M_{N}.\label{eq:stimatore}
\end{equation}

\begin{figure}
\includegraphics[scale=0.22]{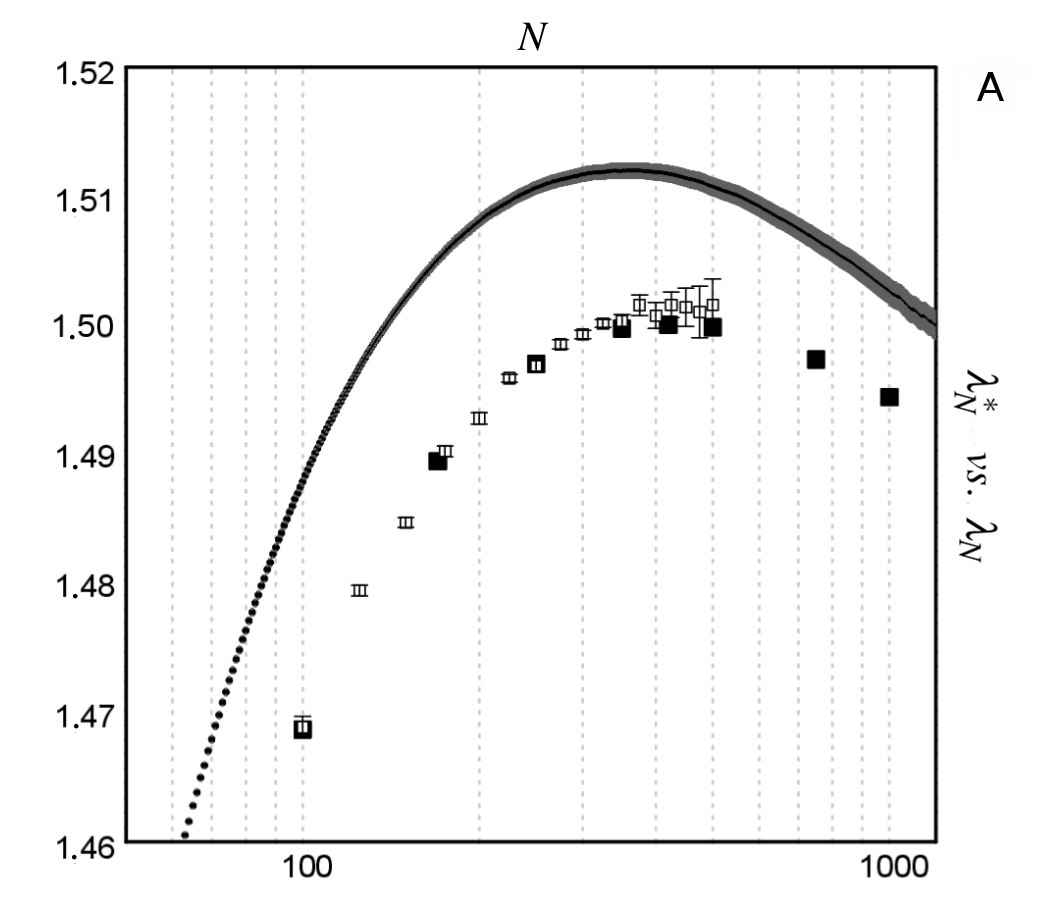}

\includegraphics[scale=0.22]{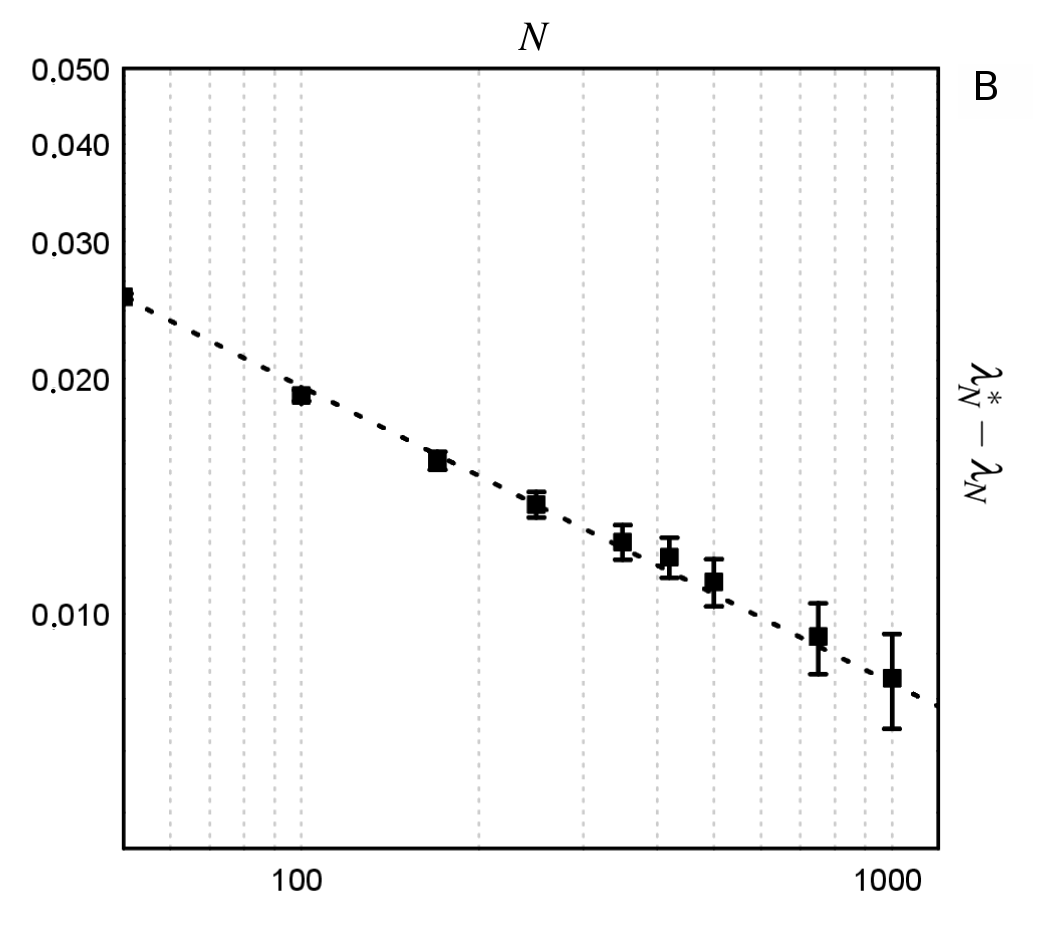}

\caption{\label{FIG6-1-1}Comparison between the critical ISAW energy from
independent PERM simulations with the estimator of Eq. (\ref{eq:stimatore})
up to $N=1000$ computed with an unbiased algorithm. In figure A,
semi-log scale, the black line is the estimator $\lambda_{N}^{*}$
with its error (standard deviation), obtained from an unbiased simulation,
while the black dots are values obtained with an independent PERM
simulation. Finally, the white squares are the intercepts at $M=0$
from linear fits of Figure \ref{FIG6}. B shows the difference $\lambda_{N}^{*}-\lambda_{N}$
between ISAW critical energy and Eq. (\ref{eq:stimatore}) in log-log
scale. The difference is fitted with a power law $K_{0}\,x^{-c}$,
with $K_{0}=0.1124\pm0.0005$ and exponent $c=-0.38\pm0.01$.}
\end{figure}

We generated SRW samples with an unbiased algorithm and compared the
above estimators with the critical energy from PERM simulations of
the ISAW. Our simulations up to $N=1000$ support the hypothesis that
the estimator $\lambda_{N}^{*}$ does eventually converge to $\lambda_{c}$
(see Figure \ref{FIG6-1-1}). We remark that such relation is due
to the fact that both the extended phase and the clustered phase scale
differently from the SRW. In higher dimensions we cannot rely on this
property because for $d>4$ the SAW is expected to scale like the
SRW. 

\section{Conclusions and Outlooks}

Concerning the form of the transition line, it is important to remark
that the conjecture in Eq. (\ref{eq:stimatore}) would open interesting
analytic possibilities. In fact, the quantity $\delta_{N}^{*}$ does
not depend on $\beta$ and all the averages are taken with respect
to the SRW measure. We expect that, apart from messy algebra, the
asymptotics of the necessary correlation functions can be computed
using the very same techniques developed by Jain and Pruitt to compute
the variance of the SRW range \cite{Huges,Jian-Pruitt,Jiain-Orey,Dvoretzky-Erdos,Den Hollander}.
This would be a nice result, since to the best of our knowledge no
exact expression is known or even conjectured for the ISAW critical
energy.

Another interesting fact is that the model can be described by a generalized
urn model. Since $L\left(\omega_{N}\right)$ can increase only if
$M\left(\omega_{N}\right)$ does not, it holds
\begin{equation}
L\left(\omega_{N+1}\right)-L\left(\omega_{N}\right)=2d\cdot\pi\left(\omega_{N+1}\right)\left\{ 1-\left[M\left(\omega_{N+1}\right)-M\left(\omega_{N}\right)\right]\right\} \label{eq:df}
\end{equation}
where we used the symbol
\begin{equation}
\pi\left(\omega_{N}\right)=\langle M\left(\omega_{N+1}\right)-M\left(\omega_{N}\right)|\omega_{N}\rangle_{0}\label{eq:df-1}
\end{equation}
to indicate the atmosphere of the chain (see \cite{frabalz}). Given
the urn kernels 
\begin{equation}
\pi_{N}^{\left(k\right)}\left(M,L\right)=\langle I\left(L\left(\omega_{N}\right)-L\left(\omega_{N-1}\right)=k\right)\rangle_{M,L}
\end{equation}
for $0\leq k\leq2d$ we conjecture that
\begin{equation}
\pi^{\left(k\right)}\left(m,\lambda\right)=\lim_{N\rightarrow\infty}\pi_{N}^{\left(k\right)}\left(\left\lfloor mN\right\rfloor ,\left\lfloor \lambda N\right\rfloor \right)
\end{equation}
exists for all considered $k$, and that it would be possible to extend
the urn techniques presented in \cite{frabalz,fra urne} to deal with
the urn model controlled by the kernels $\pi^{\left(k\right)}\left(m,\lambda\right)$.
Notice that for $k=0$ one would have
\begin{equation}
\pi_{N}^{\left(0\right)}\left(M,L\right)=\langle I\left(L\left(\omega_{N}\right)-L\left(\omega_{N-1}\right)=0\right)\rangle_{M,L}=\langle I\left(M\left(\omega_{N}\right)-M\left(\omega_{N-1}\right)=1\right)\rangle_{M,L}
\end{equation}
and that by definition must hold
\begin{equation}
1-\pi_{N}^{\left(0\right)}\left(M,L\right)=\sum_{k=1}^{2d}\pi_{N}^{\left(k\right)}\left(M,L\right).
\end{equation}

We conclude with one last remark. Due to difficulties in simulating
long chains when $m$ is close to $1$ we where unable to directly
check the behavior in this region. At first we where tempted to further
push the conjecture and guess that in the TL the critical line hits
the value $\lambda=0$ at $m=1$, but our PERM estimates seem to exclude
this simple ansatz because the observed $\lambda_{N}\left(1\right)$
is always below the value $\lambda_{c}=\lambda_{0}/\left(1-C_{3}\right)\simeq1.5238$
for which a ``linear'' critical line can pass through the point
$(m_{0},\,\lambda_{0})$, that must lie on the critical line in any
case (we estimate $\lambda_{0}\simeq1.005$ numerically
and $m_{0}=C_{3}\simeq0.3405$ from \cite{Douglas}), and then hit the
boundary $3\left(1-m\right)$ of the allowed parameter space at $m=1$
exactly.

Then, if the linear behavior of $\ell_{N}\left(m\right)$ can be really
extended in the whole $m$ range and $\lambda_{c}<\lambda_{0}/\left(1-C_{3}\right)$
this would imply the existence of a second critical value for the
intersection density, ie the $m^{*}=C_{3}\cdot(\lambda_{c}-3)/(\lambda_{c}-\lambda_{0}-3\cdot C_{3})$
at which the crossing between the critical line $\ell_{c}\left(m\right)$
and the boundary $3\left(1-m\right)$ actually happens, and after
this value the clustered phase would not be possible anymore except
for values of $\lambda$ concentrating on the boundary of the parameter
range. For or example, the conjecture would imply that no CG transition
can occur for $m<1$ in the $\Omega_{N}\left(\left\lfloor mN\right\rfloor ,\left\lfloor \left(1-m\right)N\right\rfloor \right)$
model, where the exceeding nearest neighbor pairs are forbidden. This
is likely because in a clustered phase we necessarily have a partial
saturation of the nearest neighbor sites of each monomer, and such
phase would be extremely unfavored by a small link density.

\section{Acknowledgments}

We would like to thank Giorgio Parisi (Sapienza Univeristà di Roma)
and Valerio Paladino (Amadeus IT) for interesting discussions and
suggestions. This project has received funding from the European Research
Council (ERC) under the European Union’s Horizon 2020 research and
innovation programme (grant agreement No {[}694925{]}).


\begin{thebibliography}{10}
\bibitem{Flory}P. J. Flory, \emph{Principles of Polymer Chemistry}
(Cornell University Press, 1971).

\bibitem{CG}I. Nishio, S.-T. Sun, G. Swislow and T. Tanaka, Nature
\textbf{281}, 208–209 (1979).

\bibitem{CGDNA}K. Minagawa Y. Matsuzawa K. Yoshikawa A. R. Khokhlov
and M. Doi, Biopolymers \textbf{34}, 555-558 (1994).

\bibitem{De Gennes-Scal.} P. G. de Gennes, \emph{Scaling Concepts
in Polymer Physics}, Cornell University Press (1979).

\bibitem{des Cloizeaux}J. des Cloizeaux and G. Jannink, \emph{Polymers
in solutions: Their Modelling and Structure}, Clarendon Press, Oxford
(1990).

\bibitem{Grosberg-Kuznestov}A. Y. Grosberg and D. V. Kuznetsov, Macromolecules
\textbf{25}, 1970 (1992).

\bibitem{ISAW1}P. P. Nidras, J. Phys. A: Math. Gen. \textbf{29},
7929 (1996).

\bibitem{ISAW2}M. C. Tesi, E. J. J. van Rensburgd, E. Orlandini and
S. G. Whittington, J. Phys. A: Math. Gen. \textbf{29} 2451 (1996).

\bibitem{ISAW3}M. C. Tesi, E. J. J. van Rensburg, E. Orlandini and
S. G. Whittington, J. Stat. Phys. \textbf{82}, 155–181 (1996).

\bibitem{ISAW4}S. Caracciolo, M. Gherardi, M. Papinutto and A. Pelissetto,
J. Phys. A \textbf{44}, 115004 (2011).

\bibitem{ISAW5}C.-N. Chen, Y.-H. Hsieh and C.-K. Hu, EPL \textbf{104},
20005 (2013).

\bibitem{ISAW6}N. R. Beaton, A. J. Guttmann and I. Jensen, J. Phys.
A: Math. Theo. \textbf{53}, 165002 (2020).

\bibitem{DJMODEL}C. Domb and G. S. Joyce, J. Phys. C: Solid State
Phys. \textbf{5}, 956 (1972).

\bibitem{DJMODEL2}N. Clisby, J. Phys.: Conf. Ser. \textbf{921} 012012
(2017).

\bibitem{Slade RG Theory}G. Slade, Proc. R. Soc. A, \textbf{475},
20181549 (2019).

\bibitem{Douglas}J. F. Douglas and T. Ishinabe, Phys. Rev. E \textbf{51},
1791 (1995).

\bibitem{Madras-Slade}N. Madras and G. Slade, \emph{The Self-Avoiding
Walk} (Birkhauser, Boston, 1996).

\bibitem{Clisby}N. Clisby, Phys. Rev. Lett. \textbf{104}, 055702
(2010).

\bibitem{Clisby2}N. Clisby, J. Phys. A: Math. Theor. \textbf{34},
5773 (2013).

\bibitem{Franchini}S. Franchini, Phys. Rev. E \textbf{84}, 051104
(2011).

\bibitem{frabalz}S. Franchini and R. Balzan, Phys. Rev. E \textbf{98},
042502 (2018).

\bibitem{PERM}The Pruned-Enriched Rosenbluth Method (PERM) is a classic
stochastic growth algorithm which combines the Rosenbluth-Rosenbluth
method with recursive enrichment. One starts by building instances
according to a biased distribution, then corrects for this by cloning
desired and killing undesired configurations to contain the weights
fluctuations, see \cite{Grassberger,Prellberg,Hsu-Grassberger} for
reviews and \cite{Grassberger} for a pseudocode.

\bibitem{Grassberger}P. Grassberger, Phys. Rev. E \textbf{56}, 3682
(1997).

\bibitem{Prellberg}T. Prellberg and J. Krawczyk, Phys. Rev. Lett.
\textbf{92}, 120602 (2004).

\bibitem{Hsu-Grassberger}H.-P. Hsu and P. Grassberger, J. Stat. Phys.
\textbf{144}, 597 (2011).

\bibitem{levelchoice}Although the choice $r=1$ smoothly connects
with the SRW we remark that by Eq.(\ref{eq:behavior}) the level lines
$\ell_{N}\left(m,r\right)$ will eventually converge to the critical
line for any fixed $r$.

\bibitem{logcorr}See for example the $\Omega_{N}$$\left(M\right)$
model of \cite{frabalz}, where the product measure condition is likely
to give only approximate results for any $d<\infty$ due to excluded
volume effects.

\bibitem{Spitzer}F. Spitzer, \emph{Principles of Random Walk} (Springer,
New York, 2001).

\bibitem{Huges}B. D. Hughes, \emph{Random Walks and Random Enviroments},
Vol.1 (Clarendon Press, Oxford, 1995).

\bibitem{Feller}W. Feller, \emph{An introduction to Probability Theory
and Its Applications}, Vol. 1 (Wiley, New York, 1950).

\bibitem{Jian-Pruitt}N. C. Jain and W. E. Pruitt, J. Analyse Math.
\textbf{24}, 369 (1971).

\bibitem{Jiain-Orey}N. C. Jain, S. Orey, Isr. J. Math. \textbf{6},
373 (1968).

\bibitem{Dvoretzky-Erdos}A. Dvoretzky and P. Erdos, Proc. 2nd Berkley
Symp. on Prob. and Stat., \textbf{353} (1951).

\bibitem{fra urne}S. Franchini, Stoc. Proc. Appl. \textbf{127} (2017).

\bibitem{Den Hollander}F. Den Hollander, J. Stat. Phys. \textbf{37}
(1984) 331-367.

\bibitem{Slade}D.C. Brydges and G. Slade, J. Stat. Phys. \textbf{159}
(2015) 421-667. \end{thebibliography}
\end{document}